\documentclass{elsart4-1}

\usepackage{graphicx}
\usepackage{epsfig}

\usepackage{amssymb}
\usepackage[latin1]{inputenc}

\usepackage[english,francais]{babel}

\newtheorem{e-proposition}[theorem]{Proposition}

\newtheorem{e-definition}[theorem]{Definition\rm}

\renewcommand{\theequation}{\arabic{equation}}
\def\og{\leavevmode\raise.3ex\hbox{$\scriptscriptstyle\langle\!\langle$~}}
\def\fg{\leavevmode\raise.3ex\hbox{~$\!\scriptscriptstyle\,\rangle\!\rangle$}}

\begin{document}

\centerline{Astrophysics}
\begin{frontmatter}



\selectlanguage{english}
\title{Cosmic rays around $10^{18}~$eV~: implications of contemporary measurements on the origin of the ankle feature}


\selectlanguage{english}
\author{Olivier Deligny}
\ead{deligny@ipno.in2p3.fr}

\address{IPN Orsay, 15 Rue Clemenceau\\
91406 Orsay CEDEX, France}


\medskip
\begin{center}
{\small Received *****; accepted after revision +++++}
\end{center}

\begin{abstract}
The impressive power-law decay of the energy spectrum of cosmic rays over more than thirty orders of magnitude 
in intensity and for energies ranging over eleven decades between $\simeq 10^9~$eV and $\simeq 10^{20}~$eV 
is actually dotted with small irregularities. These irregularities are highly valuable for uncovering and understanding 
the modes of production and propagation of cosmic rays. They manifest themselves through changes in the spectral 
index characterising the observed power laws. One of these irregularities, known as the \textit{ankle}, is subject to 
conflicting interpretations for many years. If contemporary observations characterising it have shed new lights, they 
are still far from being able to deliver all the story. The purpose of this contribution is to give an overview of the physics 
of cosmic rays in the energy range where the transition between Galactic and extragalactic cosmic rays is expected to 
occur, and to deliver several lines of thought about the origin of the ankle.

\vskip 0.5\baselineskip

\selectlanguage{francais}
\noindent{\bf R\'esum\'e}
\vskip 0.5\baselineskip
\noindent
{\bf Rayons cosmiques autour de $10^{18}~$eV~: implications des observations contemporaines pour l'origine de la 
cheville. }
L'impressionnante d\'ecroissance en loi de puissance du spectre en \'energie des rayons cosmiques sur plus de trente 
ordres de grandeur en intensit\'e et sur une \'echelle d'\'energie balayant onze d\'ecades entre $\simeq 10^9~$eV et 
$\simeq 10^{20}~$eV est en fait sem\'ee de petites irr\'egularit\'es riches d'enseignements quant aux modes de 
production et de propagation des rayons cosmiques. Ces irr\'egularit\'es se manifestent par des changements de l'indice 
spectral caract\'erisant les lois de puissance observ\'ees. L'une de ces irr\'egularit\'es, connue sous le nom de \textit{cheville}, 
fait l'objet d'interpr\'etations antagonistes depuis bon nombre d'ann\'ees. Si les observations contemporaines la caract\'erisant 
ont permis de mieux la cerner, elles sont n\'eanmoins encore loin de pouvoir en livrer tous les ressorts. L'objet de cette 
contribution est de donner un aper\c{c}u de la physique des rayons cosmiques dans la gamme d'\'energie propice \`a 
une transition entre rayons cosmiques d'origine Galactique et extragalactique, et de livrer plusieurs pistes de r\'eflexion sur 
l'origine de la cheville. 

\vskip 0.5\baselineskip
\noindent{\small{\it Key words~:} Cosmic rays; Iron knee; Ankle; Galactic/extragalactic transition \vskip 0.5\baselineskip
\noindent{\small{\it Mots-cl\'es~:} Rayons cosmiques~; Genou de fer~; Cheville~; Transition galactique/extragalactique}}

\end{abstract}
\end{frontmatter}


\selectlanguage{english}

\section{Introduction}
\label{sec:intro}

The ankle is a hardening of the energy spectrum of cosmic rays in the $10^{18}~$eV energy range. 
Discovered in 1963 by Linsley at the Volcano Ranch experiment~\cite{Linsley1963}, subsequent large-aperture 
experiments have nevertheless been necessary to characterise accurately this feature
~\cite{Lawrence1991,Nagano1992,Bird1993,AugerPLB2010,TA2013}. The ankle is today commonly described 
as a \textit{sharp} slope change of the spectral index of the cosmic ray intensity from $\simeq 3.3$ to $\simeq 2.7$ 
located at $\simeq 4\times 10^{18}~$eV.

Although the existence of the ankle is beyond controversy, its interpretation is still under debate. A dedicated
picture explaining this feature was already put forward by Linsley while he was reporting on this finding~:
\og There are many possible interpretations. The one we favor is that the inflection marks a crossover between
galactic and metagalactic cosmic rays\fg~\cite{Linsley1963}. In other words, this is nothing else but 
saying that the ankle may be the spectral feature marking the \textit{transition between Galactic and extragalactic 
cosmic rays}. The reasons supporting this interpretation are the object of section~\ref{sec:endgalactic}. 
Despite the fact that it has been popular for many years, this interpretation appears now in tension with contemporary 
measurements related to the mass composition of cosmic rays, the large-scale distribution of their arrival directions, 
and the energy spectrum for different mass group elements available around $10^{17}~$eV. The descriptions of 
these tensions are also the object of section~\ref{sec:endgalactic}. 

Alternatively, the ankle may be understood as the natural distortion of a proton-dominated extragalactic
spectrum due to $e^{\pm}$ pair production in the collisions with the photons of the cosmic microwave
background~\cite{Hillas1967,Blumenthal1970,Berezinsky2006,Berezinsky2004}. This requires on 
the one hand that the transition between Galactic and extragalactic cosmic rays takes place at lower energies 
and produces another spectral feature below $10^{18}~$eV~\cite{Berezinsky2006,Berezinsky2004}, and on the 
other hand that the mass composition is made of protons exclusively from the transition energy up to the highest
ones. This scenario, referred to as \textit{the dip model}, is presented in section~\ref{sec:dip}.

The Telescope Array and the Pierre Auger Observatory, which are currently running, are the two largest aperture 
experiments ever built to study cosmic rays around $10^{18}~$eV and above. Although the dip model provides 
an overall satisfactory theoretical framework in interpreting data collected at the Telescope Array, a totally different 
picture is needed in interpreting the ones collected at the Pierre Auger Observatory. This picture, though not final, is 
the object of section~\ref{sec:augerdata}.

Hence, establishing the energy at which the intensity of cosmic rays starts to dominate the intensity of Galactic 
ones is still, as of today, a fundamental question in astroparticle physics. Part of the answer lies in understanding 
the ankle, but the whole picture will clearly appear only by revealing the origin of cosmic rays down to $10^{17}~$eV. 
To this end, some signatures of different scenarios that would be interesting, though challenging, to address in the
future from an observational point of view are discussed in the final section.

\section{The end of the bulk of Galactic cosmic rays}
\label{sec:endgalactic}

Galactic cosmic rays are thought to be retained by the Galactic magnetic field as long as the size of their Larmor orbit
diameter is less than the thickness of the Galactic disk. The large confinement time of the particles in the disk is known
to make the energy spectrum observed on Earth steeper compared to the one at the sources. Since the strength of the 
magnetic field is of the order of microgauss, Galactic cosmic rays might be confined in the Galactic disk up to energies of 
$Z~10^{17}~$eV, with $Z$ the charge of the particles. Once particles are not confined anymore - which should happen
at an energy which depends on the charge $Z$, the time they spent in the disk tends to the constant free escape time due 
to the direct escape from the Galaxy. If cosmic rays up to the highest energies were produced by the same sources in the 
Galaxy responsible for the bulk of low-energy cosmic rays, the energy spectrum of each mass group element should thus 
show a prominent hardening (with a change of slope of $\simeq 1$) around each charge-dependent energy at which the 
free escape starts. Also, for energies at which particles escape freely from the Galaxy, the observed intensity 
should be naturally much stronger towards the disk compared to other directions. Due to their high level of isotropy, cosmic 
rays in excess of $\simeq 10^{18}~$eV have thus been thought to be of extragalactic origin since a long time. Because a 
hardening in the energy spectrum is a natural feature expected from the intersection of a steep component with a flatter 
one, the ankle has been widely considered as the onset in the energy spectrum marking the transition between Galactic 
and extragalactic cosmic rays~\cite{Linsley1963,Szabelski2002,Hillas2004,DeMarco2005,Wibig2005,Allard2007}. 

Meanwhile, the hypothesis that supernova remnants are the sources of the bulk of Galactic cosmic rays is considered
today as a standard paradigm. This is mainly because the intensity of cosmic rays observed on Earth can be produced 
by using $\simeq 10$\% of the energetics of these astrophysical objects~\cite{Zwicky1934}; and because the 
diffusive shock acceleration has been shown to be a mechanism able to convert kinetic energy of the expanding
supernova blast wave into accelerated particles~\cite{Bell1978a,Bell1978b,Blanford1987}. In this case, the 
rigidity-dependent maximum acceleration energy for Galactic cosmic rays is expected to be of the order of $10^{17}~$eV 
for iron nuclei~\cite{Berezhko2007}. This makes it difficult to reach smoothly the ankle energy.

Besides this theoretical difficulty, it is worth stressing the fine-tuning required to support the hypothesis that the transition
occurs at the ankle energy. Pioneer measurements were sparse, and in particular the measurement of the ankle was not very 
accurate. Only a smooth change of slope could be observed over a wide energy range in overall agreement with the fact
that a transition from a steep spectral index ($\gamma\simeq -3.3$) to a flatter one ($\gamma\simeq -2.7$) ranges smoothly 
over about one decade in energy. In contrast to this natural smooth hardening, an impressive feature of contemporary 
measurements of the ankle is its \textit{sharpness}. To reproduce such a feature, the end of the steep component has to 
\textit{sharply cut off} at the exact same energy at which the flatter component becomes predominant, so that the overall 
energy spectrum appears as a continuous function. \textit{This involves fine-tunings}, and calls into question the natural 
aspect of the ankle as the transition. 

\begin{figure}[!t]
  \centering
  \includegraphics[width=7.5cm]{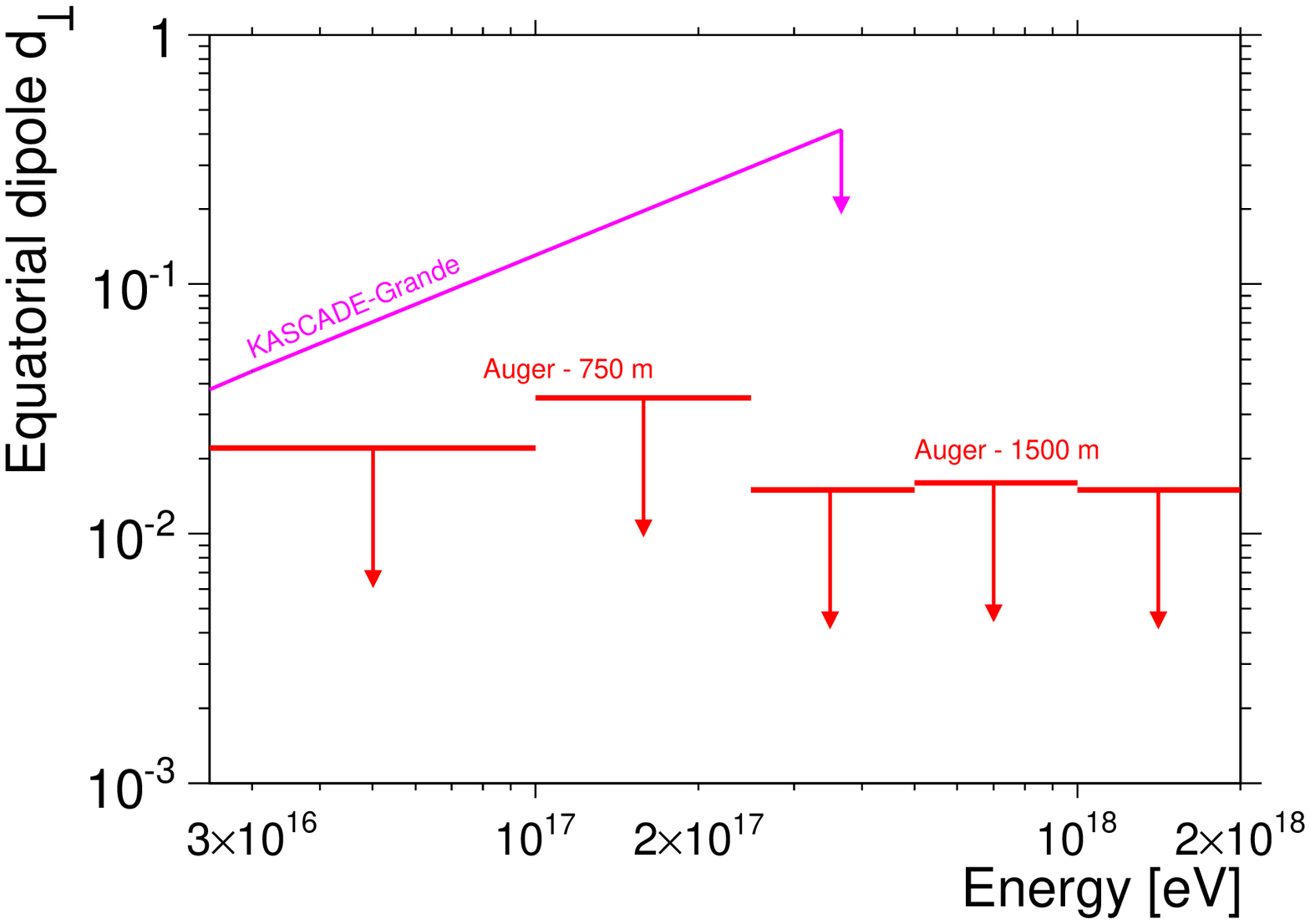}
  \includegraphics[width=7.5cm]{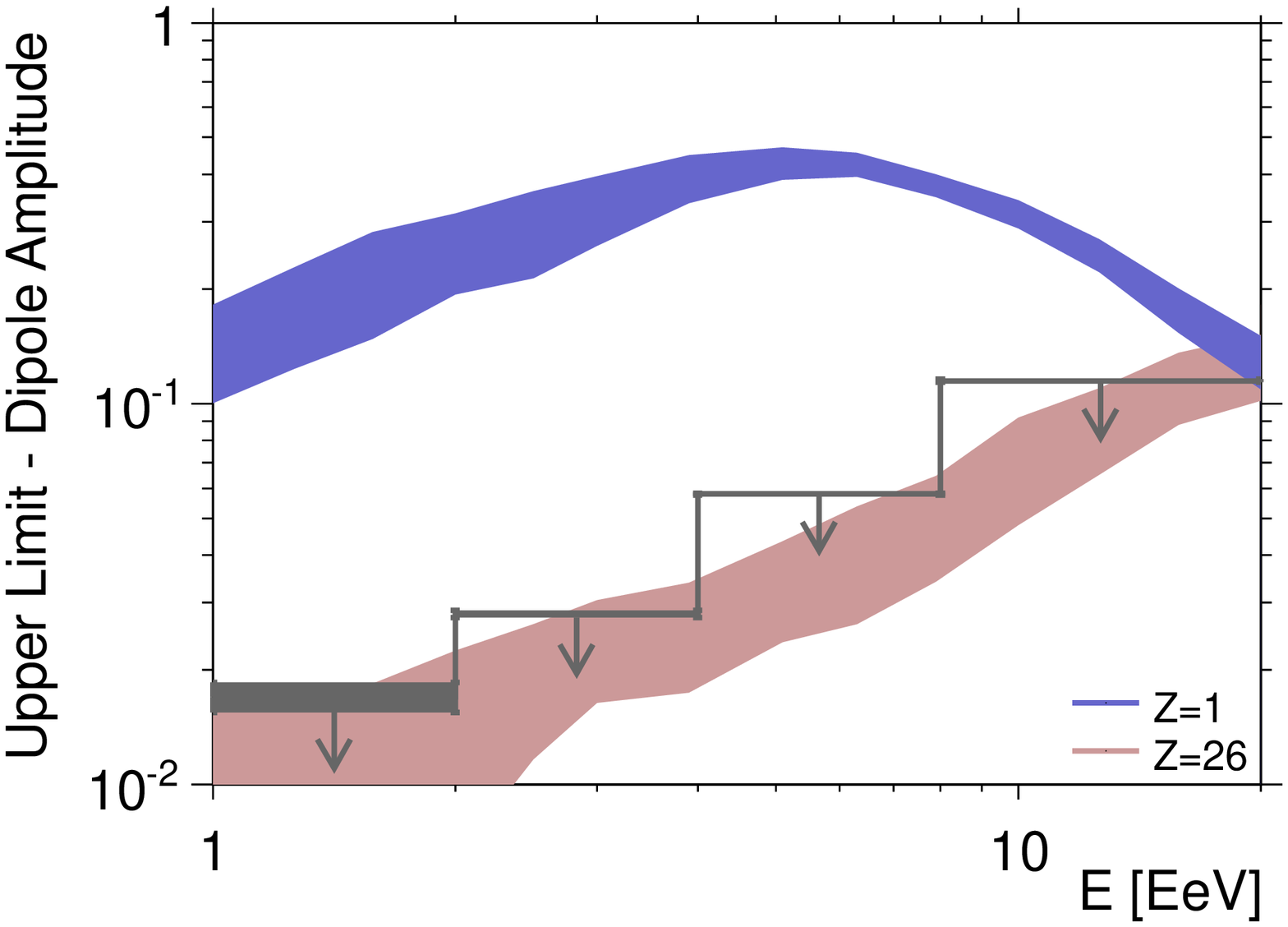}
  \caption{Left~: 99\%confidence level upper limits on the equatorial component (in equatorial coordinates) of a dipole 
  as a function of the energy. In red are the limits obtained over the full energy range of the Auger observatory~\cite{Sidelnik2013}.
  Are also shown in magenta the limits from KASCADE-Grande~\cite{Over2007}. 
  Right~: 99\%confidence level upper limits on a dipole amplitude on the celestial sphere as a function of the energy above
  $10^{18}$~eV. Some generic anisotropy expectations from stationary Galactic sources distributed in the disk are also shown 
  for two extreme cosmic-ray compositions. From~\cite{AugerApJL2013,AugerApJS2012}.}
  \label{fig:upper_limits_anis}
\end{figure}

Recent observational results provide interesting additional constraints on the origin of the ankle. One of these results is the 
high level of isotropy observed at the Pierre Auger Observatory~\cite{AugerAPP2011,AugerApJL2013,AugerApJS2012}. Using 
the large amount of data around $10^{18}$~eV collected at this observatory, and throughout the amplitudes measured in 
different energy ranges, no significant departure from isotropy has been observed be it in terms of first harmonic modulations 
in the right ascension distributions (left panel of figure~\ref{fig:upper_limits_anis} lastly updated in~\cite{Sidelnik2013}), or 
of modulations following a dipole pattern on the celestial sphere (right panel). The derived upper limits are stringent enough to 
challenge some Galactic scenarios. Indeed, large-scale anisotropies in the distribution of arrival directions is a natural signature 
of the escape of cosmic rays from the Galaxy. For stationary sources densely distributed in the disk and emitting particles in all 
directions, estimates based on Monte-Carlo simulations of particles propagating in the Galactic magnetic field show that the 
dipole amplitudes expected for protons largely stand above the allowed limits~\cite{AugerApJL2013,AugerApJS2012} (see also~\cite{Giacinti2012}). To respect the dipole limits below the ankle energy, the fraction of protons should not exceed 
$\simeq 10$\% of the cosmic-ray composition. This is particularly interesting in the view of the indications for the dominance of 
a light component around $10^{18}$~eV (see below). These constraints on this Galactic scenario are also supported by the 
absence of any detectable point-like source above $10^{18}$~eV that would be indicative of a flux of neutrons produced by 
high-energy protons through charge exchange interactions in the source environment~\cite{AugerApJ2012}. 

\begin{figure}[!t]
  \centering
  \includegraphics[width=7.5cm]{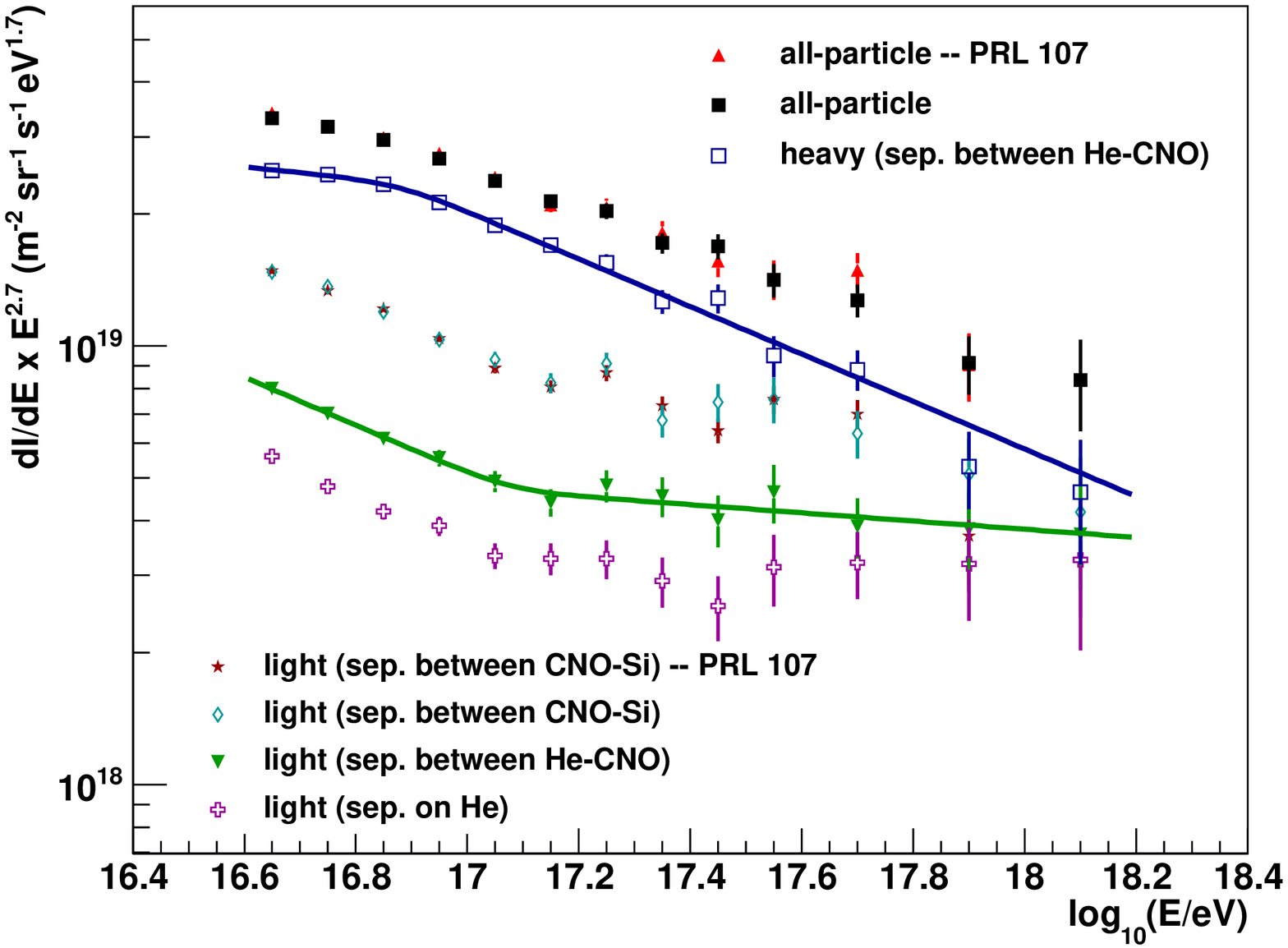}
  \includegraphics[width=8.0cm]{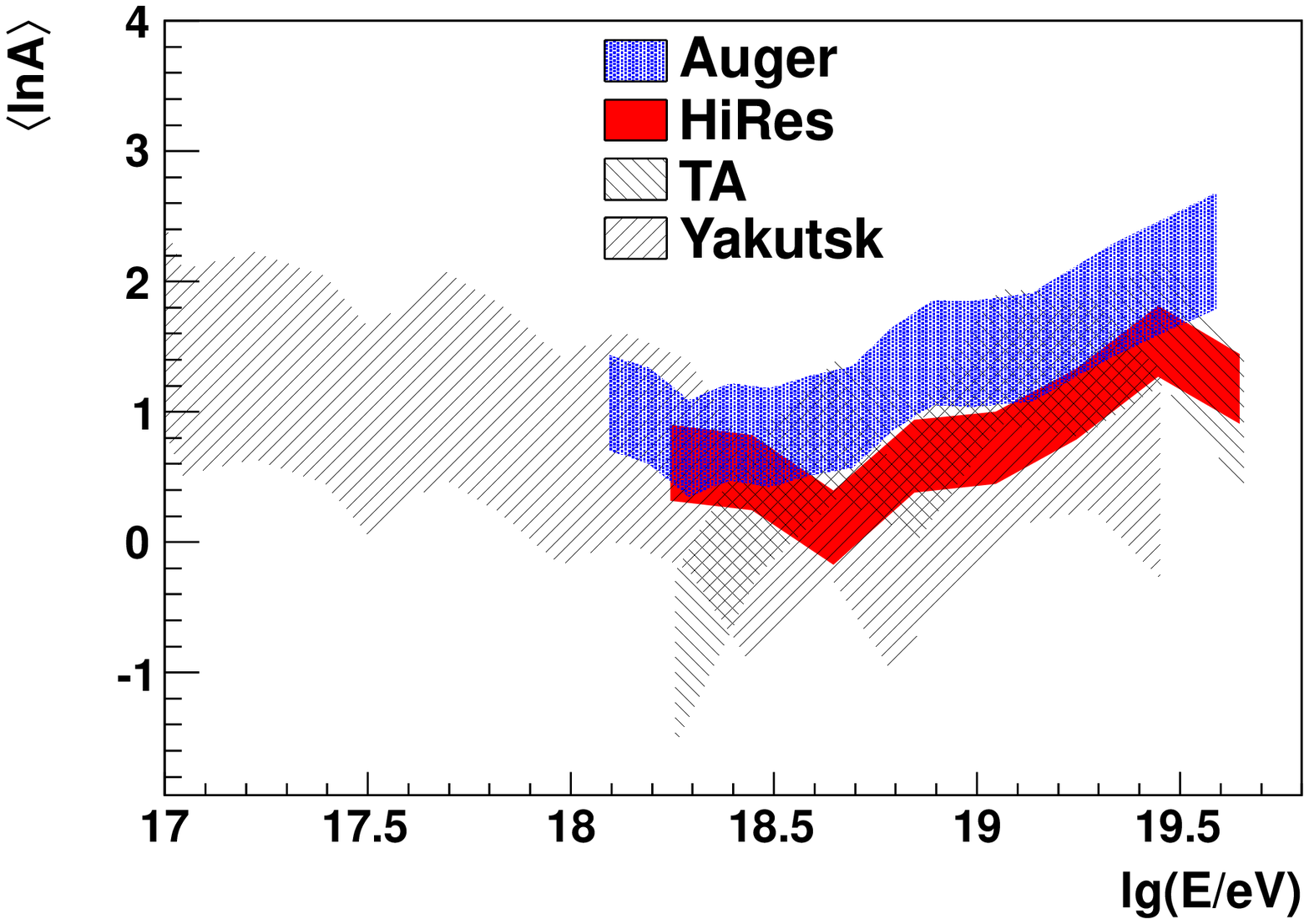}
  \caption{Left~: All-particle spectrum as measured at the KASCADE-Grande experiment, together with the separate energy 
  spectrum for light and heavy elements~\cite{KASCADEGrande2013}. The separation is based on a distinction selecting 
  preferentially protons and He nuclei in the light element sample, and selecting preferentially CNO, Si and Fe nuclei in the
  heavy element sample. Right~: Averaged composition in terms of the logarithmic mass inferred from measurements of the 
  mean atmospheric depth $\left<X_{\mathrm{max}}\right>$ at the Auger, HiRes, Telescope Array and Yakutsk experiments
  (from~\cite{Barcikowski2013}).}
  \label{fig:xmax_spectrumkg}
\end{figure}

The second measurement of interest is the energy spectrum between $\simeq 10^{17}$ and few $\simeq 10^{18}~$eV,
which has been recently achieved with unprecedented precision. An important spectral feature, namely the \textit{iron knee}, 
has been observed at $\simeq 10^{17}~$eV~\cite{KASCADEGrande2011,Tunka2012}, energy at which the mass composition 
is dominated by iron-nuclei elements. This name, corresponding to a softening of the all-particle energy spectrum, is due to 
the similarity to the knee feature located at an energy about twenty six times lower as measured by the KASCADE 
experiment~\cite{KASCADE2003,KASCADE2005} and others, \textit{e.g.} the EAS-TOP one~\cite{EASTOP2004a,EASTOP2004b}. 
This observation is thus supporting the paradigm of rigidity-dependent acceleration at the sources. In addition, a recent milestone 
was achieved by measuring separately the energy spectrum of light and heavy elements above 
$\simeq 10^{17}~$eV~\cite{KASCADEGrande2013}. It turns out from this measurement that, as shown in the left panel of 
figure~\ref{fig:xmax_spectrumkg}, the softening feature can be unambiguously attributed to the heavier elements on the one 
hand, and that a \textit{hardening} is observed at about the same energy in the energy spectrum of the \textit{light} elements 
on the other hand. Note that the separation is based on a distinction selecting preferentially protons and He nuclei in the light 
element sample, and selecting preferentially CNO, Si and Fe nuclei in the heavy element sample.

The third relevant piece of information is provided by the observation of the chemical composition of cosmic rays. The 
best suited tool to infer the mass composition in this energy range is based on the measurements of the atmospheric 
depth $X_{\mathrm{max}}$ at which showers reach their maximum of development. Around $10^{18}~$eV, there 
are indications that the cosmic-ray composition includes a significant \textit{light} component~\cite{AugerPRL2010,Abbasi2010,Jui2011,Berezhko2012}, as summarised in the right panel of 
figure~\ref{fig:xmax_spectrumkg} where the mean atmospheric depth $\left<X_{\mathrm{max}}\right>$ is converted 
into the average logarithmic mass $\left<\mathrm{ln}A\right>$ by means of a widely-used interaction 
model~\cite{Barcikowski2013}. In this plot, a proton-dominated composition thus results in $\left<\mathrm{ln}A=0\right>$
while an iron-dominated composition would result in $\left<\mathrm{ln}A\simeq4\right>$. The shaded regions correspond 
to the systematic uncertainties range.

In the context of the ankle marking the transition, the highest energy particles reaching the ankle energy are generally 
considered to be iron nuclei. This is because iron nuclei can obviously be confined in the Galactic disk up to higher energies 
than lighter ones. The results shown in figure~\ref{fig:xmax_spectrumkg} provide however clear evidences that the 
contribution of iron nuclei to the total intensity, if any, would only be subdominant. In a recent and comprehensive investigation 
of the supernova remnants paradigm as the sources of Galactic cosmic rays, it has been shown that both the energy spectrum 
and the chemical composition up to $10^{17}~$eV can be reproduced in a satisfactory way for a standard diffusion in the Galaxy, 
accounting for the spatial and temporal stochastic distributions of supernova remnants~\cite{Amato2011}. Since the energy 
spectrum gets cut off exponentially at the sources above the maximum acceleration energy, an \textit{additional} component 
of light cosmic rays is then required to fill the overall energy spectrum as observed above $10^{17}~$eV and to reproduce 
chemical composition measurements around $10^{18}~$eV. Measurements shown in both panels of 
figure~\ref{fig:xmax_spectrumkg} provide good support to this scenario.

Overall, the hypothesis that the ankle marks the transition between the end of Galactic cosmic rays and extragalactic ones is 
in clear tension with contemporary observational results. The iron knee is most likely the feature marking the end of the bulk 
of Galactic cosmic rays. Alternative interpretations of the ankle are thus needed.

\section{The ankle as a signature of extragalactic protons}
\label{sec:dip}

An alternative interpretation of the ankle is possible in terms of energy spectrum distortions caused by fundamental 
particle physics processes during the propagation of ultra-high energy cosmic rays if they are composed quasi-exclusively 
of extragalactic protons~\cite{Berezinsky2006,Berezinsky2004}. Indeed, while propagating in the Universe through the 
cosmic microwave background, the most energetic cosmic-ray protons would have to undergo two interactions: photopion 
production $p+\gamma_{\mathrm{cmb}}\rightarrow N+\pi$ and $e^+e^-$ production 
$p+\gamma_{\mathrm{cmb}}\rightarrow p+e^++e^-$. The energy losses caused by these interactions would imprint 
clear signatures on the energy spectrum of extragalactic protons produced by sources distributed throughout the Universe. 

The first interaction results in a sharp steepening above $\simeq 6~10^{19}~$eV called the 
\textit{GZK cutoff}~\cite{Greisen1966,Zatsepin1966}. A comprehensive description of the GZK cutoff and of the current 
related issues is given elsewhere in this review~\cite{Harari2014}. The second interaction is, on the other hand, relevant
for providing an interpretation of the ankle in terms of a propagation effect. Provided that the initial energy of protons is 
high enough to produce electron-positron pairs through collisions with the photons of the cosmic microwave background,
the energy spectrum observed on Earth is expected to show the impact of these interactions by a steepening in an energy 
range where the protons with degraded energies pile up, followed by a return to the injected slope at higher energies. In
other words, the effect of the energy losses by pair production results in a shallow deepening called \textit{dip}, giving a 
natural way to induce an ankle spectral feature~\cite{Hillas1967,Blumenthal1970,Berezinsky2006,Berezinsky2004}. Given 
the distribution in energy of the photons of the cosmic microwave background~\footnote{The impact of the extragalactic 
background light can be neglected for the propagation of ultra-high energy protons.}, precise
calculations~\cite{Berezinsky2006,Berezinsky2004} show that the ankle is expected at an energy compatible with the one 
measured by contemporary experiments. This is a remarkable argument in favor of this scenario. Note that the steepening 
can be amplified for time-evolving sources stronger in the past than at present day. This is because a large part of the proton 
intensity observed now on Earth has passed through a photon field of higher density and higher energy per photon, which 
amplifies to some extent the effect of the energy losses by pair production on the resulting energy spectrum. 

\begin{figure}[!t]
  \centering
  \includegraphics[width=10cm]{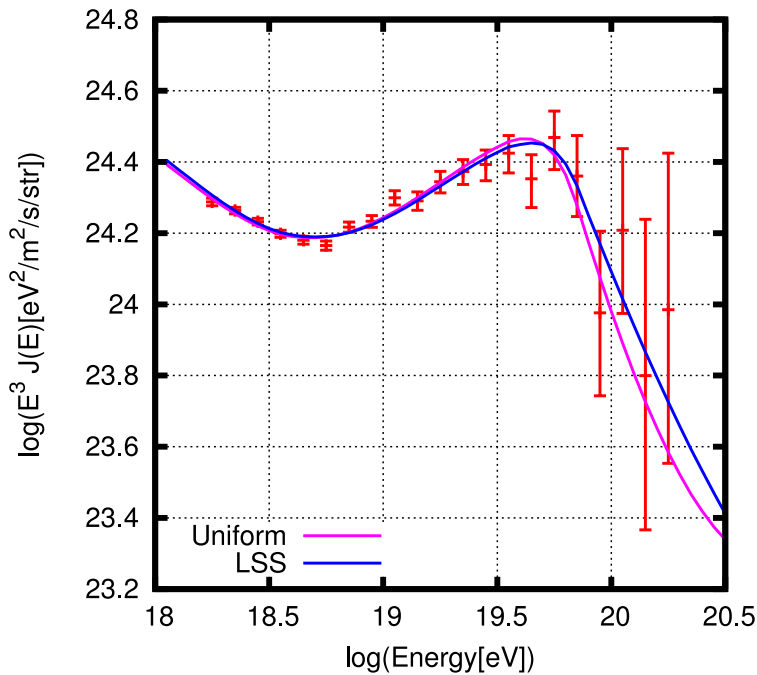}
  \caption{The energy spectrum observed at the Telescope Array interpreted in the framework of the dip model. From~\cite{Kido2013}.}
  \label{fig:spectrum_ta}
\end{figure}

In this framework, the transition between Galactic and extragalactic cosmic rays must be completed at $10^{18}~$eV
and has consequently to occur at a lower energy than the ankle one. The observation (though with limited statistics) of 
a second knee feature around $5~10^{17}~$eV by the HiRes experiment could be interpreted as the spectral feature 
marking the transition~\cite{Berezinsky2006,Berezinsky2004}. Note that the second knee is \textit{not} the iron knee. 
This name was adopted at a time where the iron knee had not been yet observed. However, this spectral feature has not 
been yet confirmed by subsequent experiments collecting much larger statistics on the one hand, and a fine-tuning is 
required to sharply cut off the extragalactic component below the second knee energy on the other hand. Instead of 
producing a sharp break in the energy spectrum, the transition might occur in a smooth and progressive way. A magnetic
horizon effect provides an interesting mechanism to achieve such a transition~\cite{Lemoine2005}. Indeed, the magnetic 
field permeating the intergalactic space, though small, may be sufficient to reduce the horizon of cosmic rays below some 
energy and thus suppress the intensity of extragalactic protons below that energy. Note also that such a low-energy cutoff 
may be intrinsic to the sources if they harbor relativistic shocks. For an exponential cutoff located close to $10^{17}~$eV, 
the all-particle energy spectrum can thus be reproduced in the context of the dip scenario, the transition taking place 
progressively below $10^{18}~$eV~\cite{Amato2011}. This is also in qualitative agreement with the energy spectrum of 
the light elements measured at the KASCADE-Grande experiment. 

The mass composition inferred from the data collected at the HiRes and Telescope Array experiments is compatible with
the hypothesis of pure protons up to the highest energies. The dip model thus provides a relevant framework for describing 
these data. Using Monte-Carlo simulations of the one-dimensional trajectories (that is, neglecting the effect of the magnetic 
fields) accounting for all energy losses and cosmological evolutions of interest, and using a maximum likelihood analysis to 
compare the observed and expected number of events in each energy bin, the energy spectrum measured at the Telescope 
Array has recently been confronted to the dip model~\cite{Kido2013}. The best fit, obtained for a spectral index at the sources 
of 2.36 and a parameter $m=4.5$ describing the source number evolution with redshift $z$ (in $(1+z)^m$), is displayed in
figure~\ref{fig:spectrum_ta} for a uniform distribution of sources and for sources following the large-scale structure of matter 
in the Universe. It turns out that, within the systematic uncertainties on the energy scale, an overall excellent agreement is 
observed. In particular, the ankle is remarkably reproduced by the dip feature. On the other hand, more statistics are needed 
to settle whether the high-energy end of the spectrum can be described by the steepening expected from the aforementioned
pion-production interactions or not.

Although the dip model provides a sufficient framework to describe the data collected at the Telescope Array, it is in tension with 
the mass composition inferred from measurements performed at the Pierre Auger Observatory. The description of these data 
requires a totally different framework, which is the object of the next section.

\section{The Auger data~: an appeal in favor of an additional component}
\label{sec:augerdata}

Provided that there is no unexpected change of the properties of the hadronic interactions describing the first steps in the shower
developments in the atmosphere at ultra-high energies, the mean depth of shower maximum and the corresponding fluctuations 
measured at the Pierre Auger Observatory provide evidences for a transition from light elements at $\simeq 10^{18}~$eV to heavier 
ones up to $\simeq 3~10^{19}~$eV~\cite{AugerPRL2010,AugerJCAP2013}. This only observation breaks down the interpretation of 
Auger data in terms of the dip model.

\begin{figure}[!t]
  \centering
  \includegraphics[width=7.5cm]{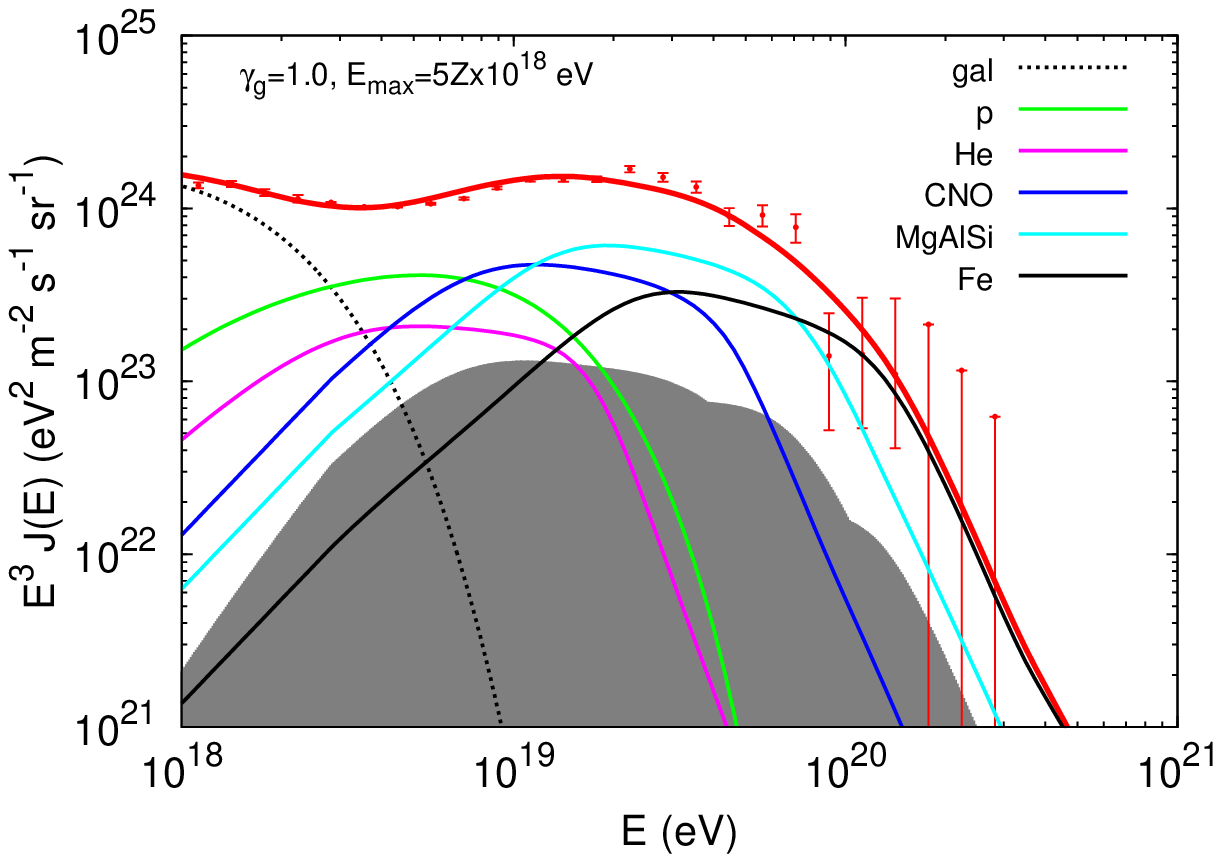}
  \includegraphics[width=7.5cm]{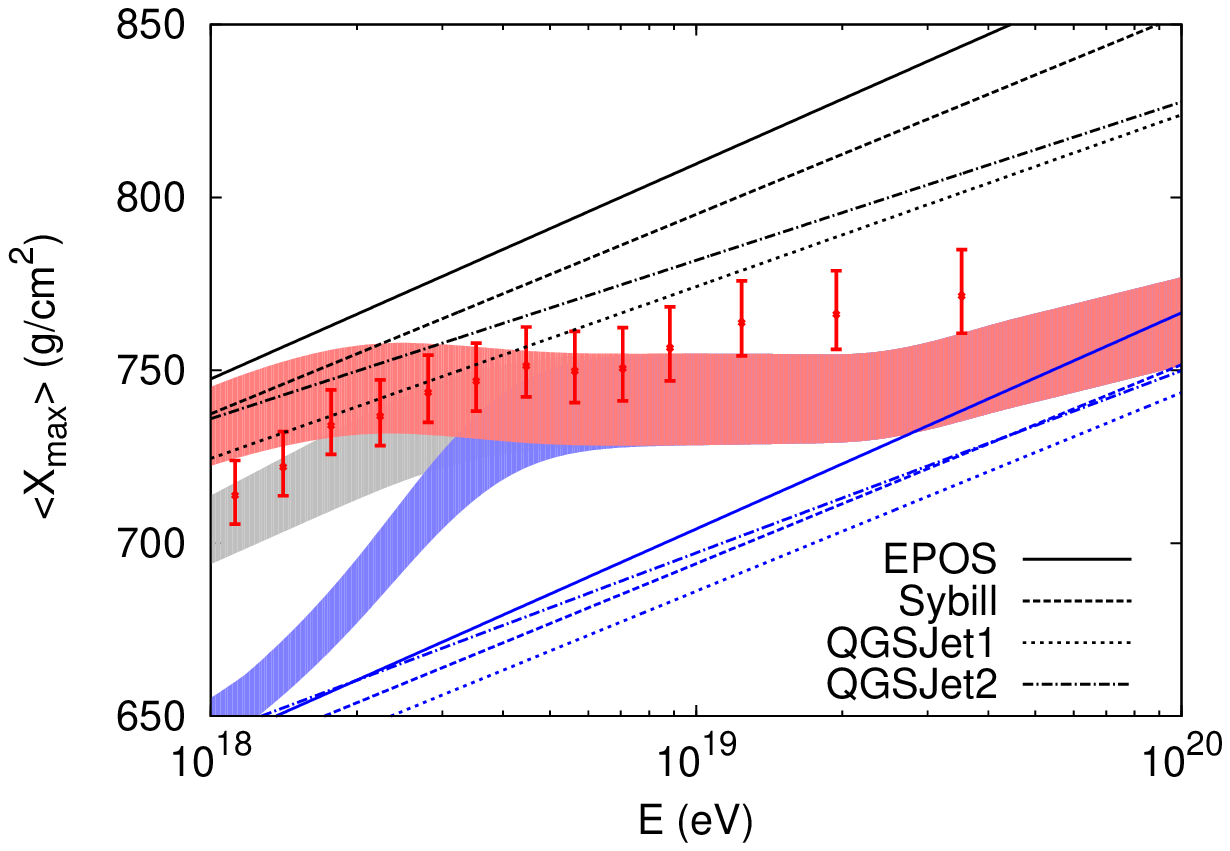}
  \caption{Left~: Intensity of different mass group elements of extragalactic origin required to reproduce simultaneously the
  energy spectrum and the chemical composition measured at the Pierre Auger Observatory, within a scenario of 
  rigidity-dependent acceleration at the sources. Right~: Results in terms of the mean value of the depth of shower
  maximum (observable sensitive to the chemical composition). From~\cite{Aloisio2014}.}
  \label{fig:auger_fit}
\end{figure}
To reproduce the energy spectrum \textit{and} the related-$X_{\mathrm{max}}$ measurements performed at the Pierre
Auger Observatory, a rigidity-dependent scenario similar to the one observed between the knee and the iron knee energies
is required~\cite{Allard2008,Taylor2010,Aloisio2014}. In this scenario, the maximum energy of the accelerators scales with
the charge number $Z$ for each element. A transition from light to heavier elements is then a natural consequence, the 
components of heavier elements being shifted in energy by the charge $Z$. Applied to the Auger data through a simultaneous
fit of the energy spectrum and of the $X_{\mathrm{max}}$ measurements, an important outcome is that the cutoff energy
for protons is found around the ankle energy~\cite{Aloisio2014}. Depending on the exact hypotheses formulated on the species 
abundances at the sources, the upper end of the energy spectrum might be due to the cutoff of the source spectrum rather than 
caused by energy losses - or to a mix between these two mechanisms.

A comprehensive discussion of these results is out of the scope of this contribution. An important outcome of this framework 
relevant for understanding the origin of the ankle is that there is no way to reproduce both the energy spectrum and the chemical
composition below $\simeq 5~10^{18}~$eV without \textit{introducing a new component mainly composed of light elements} to 
fill the gap of the all-particle energy spectrum between $\simeq 10^{17}~$eV and $\simeq 4~10^{18}~$eV, and to reproduce the
measurements related to mass composition. An illustration of this new component is given in the left panel of figure~\ref{fig:auger_fit} 
where the intensities of the different mass group elements of extragalactic origin are shown together with the additional light 
component below $\simeq 5~10^{18}~$eV, while the resulting chemical composition is shown in the right panel. The new component, 
falling off at the ankle energy, appears thus to be the high energy counterpart of the light elements responsible for the ankle-like 
feature around $10^{17}~$eV (see section~\ref{sec:endgalactic}). As such, it is \textit{not} a simple continuation of the bulk of 
Galactic cosmic rays of lower energies. 

All in all, the ankle may get back the status of a transition, but not between the bulk of Galactic cosmic rays and extragalactic ones.
However, the origin of the new component remains to be understood, as well as the mechanism which induces the sharp cutoff at 
$\simeq 4~10^{18}~$eV required to reproduce the ankle feature.

\section{Outlook}
\label{sec:discussion}

\begin{figure}[!t]
  \centering
  \includegraphics[width=10cm]{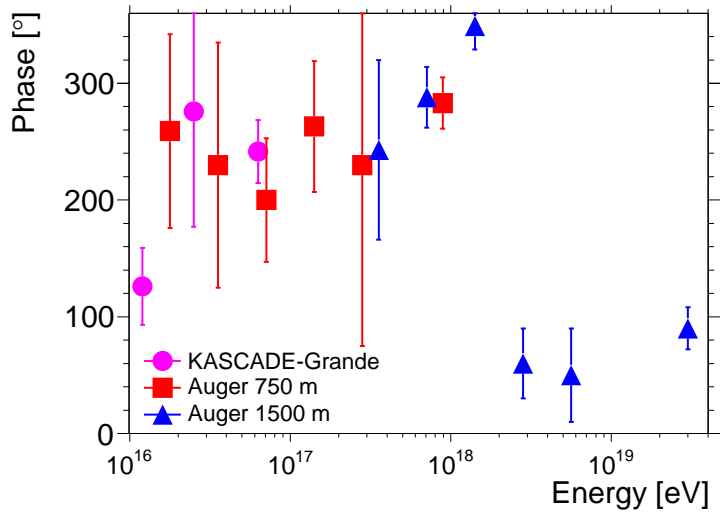}
  \caption{Phase of the first harmonic modulation in right ascension as a function of energy, from~\cite{Sidelnik2013,Curcio2013}.}
  \label{fig:phases}
\end{figure}

Despite the fact that the ankle was discovered in the early sixties, the origin of the ankle in the energy 
spectrum of cosmic rays is still a hot topic. Given the radically different interpretations of this feature which are 
possible from the data collected at the two contemporary large-aperture observatories  - namely the Telescope Array 
and the Pierre Auger Observatory, uncovering and understanding the sources of systematic uncertainties in the 
respective measurements of the energy spectrum and of the atmospheric depths where the showers reach the 
maximum of their development is of central importance. 

Large-scale anisotropies are another part of the story which might help us understanding what is going on.
Although no significant amplitude has been detected so far between $10^{17}$ and $10^{18}~$eV, 
it is to be noted that the phase measurements in adjacent energy intervals  do not seem to be randomly distributed 
but rather to follow a constant behaviour as seen in figure~\ref{fig:phases}. This is potentially indicative of a genuine 
signal, because with a real underlying anisotropy, a consistency of the phase measurements in ordered energy 
intervals is indeed expected to be revealed with a smaller number of events than needed to detect the amplitude 
with high statistical significance~\cite{Edge1978,AugerAPP2011}. Whether this consistency provides a real evidence 
for anisotropy is currently being tested at the Pierre Auger Observatory, and will be established with 99\% confidence level 
if the ongoing prescribed test is successful~\cite{AugerApJS2012,Sidelnik2013}.

\begin{figure}[!t]
  \centering
  \includegraphics[width=10cm]{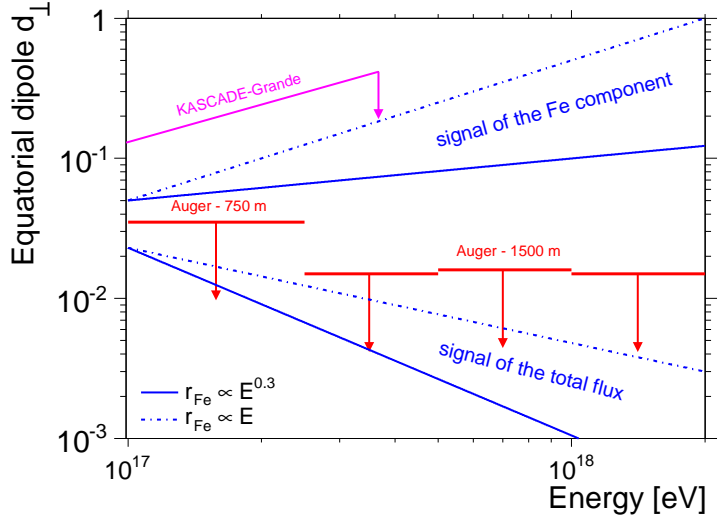}
  \caption{Illustration of a mechanism authorising a large first-harmonic amplitude for the end of the Galactic 
  component - see text.}
  \label{fig:toyanis}
\end{figure}
The putative dipole-like signal responsible for the phase alignment is required to have an amplitude rather low, being 
below the percent level to meet the conditions fixed by the upper limits obtained from the whole population of cosmic 
rays detected between $10^{17}$ and $10^{18}~$eV. Although hardly predictable in a quantitative way given the 
unknown source distributions in space and time and the difficulty to model the propagation of particles around 
$10^{18}~$eV in the intervening magnetic fields known only approximately, such an amplitude seems too low 
to be naturally explained by a Galactic scenario. On the other hand, isotropy down to a higher level is a robust
expectation in this energy range for an extragalactic scenario. A positive outcome of the prescribed test would thus
bring new pieces of information to the global puzzle. Different possibilities are discussed below.

An interesting possibility to explain such a low global level of anisotropy could be that the component of 
heavy elements marking the end of the Galactic cosmic rays is slowly extinguishing from $10^{17}$ up to 
$\simeq 10^{18}~$eV, and does show important large-scale anisotropies in its distribution of arrival 
directions. Accounting for both the distribution of supernova remnants in the Galaxy and the random distribution 
in space and time of the nearest remnants, and considering the propagation of cosmic rays in a purely turbulent 
magnetic field, recent results \textit{extrapolated} above $10^{17}~$eV suggest as a sensible possibility an 
amplitude of the equatorial component of the dipole at the level of a few percents~\cite{Amato2012}. 
At these energies, the expected amplitudes could be amplified by drift motions induced by the regular component of 
the magnetic field~\cite{Ptuskin1993,Candia2003}. The anisotropy of this subdominant component of iron nuclei 
could be diluted in the almost isotropic predominant component of light elements - possibly extragalactic 
protons - so that the anisotropy of the all-particle component could satisfy the upper limits obtained between 
$10^{17}~$eV and $10^{18}~$eV. To illustrate this scenario, let's consider the total flux 
$\Phi_{\mathrm{tot}}(E,\mathbf{n})$ as the sum of a dominant isotropic component $\Phi_X(E)$
and of a subdominant anisotropic component $\Phi_{\mathrm{Fe}}(E,\mathbf{n})$ of iron-nuclei elements~:
\begin{equation}
\Phi_{\mathrm{tot}}(E,\mathbf{n})=\Phi_X^0E^{-\gamma_X}+\Phi_{\mathrm{Fe}}^0E^{-\gamma_{\mathrm{Fe}}}(1+r_{\mathrm{Fe}}(E)~\mathbf{d}\cdot\mathbf{n}).
\end{equation}
The anisotropy of the iron component is here characterised by an energy-dependent amplitude $r_{\mathrm{Fe}}(E)$
and a direction $\mathbf{d}$. The anisotropy amplitude of the total flux is then diluted as follows:
\begin{equation}
r_{\mathrm{tot}}(E)=\frac{\Phi_{\mathrm{Fe}}^0E^{-\gamma_{\mathrm{Fe}}}}{\Phi_X^0E^{-\gamma_X}+\Phi_{\mathrm{Fe}}^0E^{-\gamma_{\mathrm{Fe}}}}r_{\mathrm{Fe}}(E).
\end{equation}
For reasonable choices of the normalisation parameters and spectral indexes such that $\gamma_{\mathrm{Fe}}$
is larger than $\gamma_{\mathrm{X}}$, a \textit{decreasing} amplitude $r_{\mathrm{tot}}(E)$ with energy can be naturally 
obtained in this scenario even for an anisotropy of the iron component increasing with energy. This is 
illustrated in figure~\ref{fig:toyanis} for two \textit{toy} energy evolutions of the anisotropy inspired from diffusion-dominated 
($r_{\mathrm{Fe}} \propto E^{0.3}$) or drift-dominated ($r_{\mathrm{Fe}} \propto E$) propagation of cosmic rays in the 
Galactic magnetic field. Although lots of fine-tunings are involved in this simplistic illustration, the global picture depicted in this 
plot may not be too irrealistic.

Another possibility to mention is that the anisotropy of the iron component is not diluted, but almost 
\textit{compensated} by an anisotropy of the same order imprinted in the arrival directions of the component of 
light elements. Since the anisotropies discussed here are mainly described in terms of dipoles, this condition can 
be met by considering that the vectors characterising the dipoles of each component have an amplitude of the same 
order but an almost \textit{opposite} direction. This scenario provides a mechanism to reduce significantly the 
amplitude of the vector describing the arrival directions of the whole population of cosmic rays - which is 
observationally the only one at reach so far. This could relax to some extent the constraints on a Galactic origin 
of the new component. Along these lines, it is interesting to note the change of phase observed around $10^{18}~$eV 
in figure~\ref{fig:phases} between $\simeq 260^\circ$ and $\simeq 80^\circ$ in right ascension. 

Alternatively, if the iron-nuclei component is strongly cut off right after $10^{17}~$eV, the putative anisotropy
would pertain to the component containing light elements. An anisotropy at the percent 
level in the $10^{18}~$eV energy range would challenge an extragalactic origin of this new component for 
equal-intensity sources spread over the Universe. Few strong local sources would be necessary to produce a strong 
density gradient of cosmic rays in the neighborhood of the Galaxy, so that an anisotropy could be observed on the 
condition that the pattern at the entrance of the Galaxy is not totally washed out by the Galactic magnetic field. 
On the other hand, a Galactic origin of this component was already shown to be in tension with the larger anisotropy 
amplitudes expected from stationary sources densely distributed predominantly in the disk and emitting in all 
directions - unless the strength of the Galactic magnetic field is much higher than currently known and/or much more 
extended perpendicularly to the disk. It is to be noted, however, that approximate anisotropy estimates from 
\textit{strongly intermittent} Galactic sources (\textit{e.g.} long gamma-ray bursts) show that the upper limits might 
be respected~\cite{Eichler2011,Kumar2013}. In this scenario, the observation of this extra-component of Galactic origin 
would be indicative of the remnant of an old and mostly isotropised population of cosmic rays from strongly intermittent 
sources \textit{different} from the ones producing the bulk of Galactic cosmic rays at lower energies. Large fluctuations 
in the intensity of the energy spectrum and in the anisotropies around $10^{18}~$eV are then intrinsic to this scenario. 
Current observations would not be representative of the usual ones and would be possible only during rare time periods. 
This could have some anthropic implications, which are however out of the scope of this report. 

That the ankle feature has not yet revealed all its secrets is indubitable. The end of the story will come from a 
comprehensive understanding of the origin of the different components observed between the iron knee and the 
ankle energies. An important step forward from an observational point of view would be to get at energy spectrum 
and anisotropy measurements for different mass group elements. This requires investing many experimental efforts 
to identify the primary mass on an event-by-event basis. A fair amount of exciting work remains.

\section*{Acknowledgements}
The progresses on the ankle origin summarised in this contribution would not have been possible without the strong 
commitment from the members of the Pierre Auger collaboration and the technical and administrative staff members
in Malarg\"{u}e. I am grateful to them.


\end{document}